# DEVELOPMENT OF A SMART MODULATOR AND EFFICIENCY EVALUATION OF 500-GEV e+e- C-BAND LINEAR COLLIDER

J. S. Oh, J. S. Bak, M. H. Cho, W. Namkung, PAL, Korea
K. H. Chung, SNU, Korea
T. Shintake, H. Matsumoto, KEK, Japan

*Abstract*

An e+e- linear collider at 500-GeV C.M. (center of mass) has been proposed as a future accelerator. The C-band linear collider has more than a thousand klystrons and matching modulators. The linear colliders require smart modulators with high reliability, compact and simple configuration, and good power efficiency. This study suggests solutions to problems expected in the future linear colliders. A C-band smart modulator has been developed using a constant-current inverter power supply. The smart modulator provides 350-kV pulses of the pulse-to-pulse stability of 0.35%, the pulse-width of 4.35 µs, and the repetition rate of 50 Hz. We also investigated the power efficiency for the C-band scheme. A pulse efficiency of 75% can be achieved by an optimized pulse transformer and the RF phase modulation. With pulsed klystrons of 60% power efficiency, it is possible to achieve the RF system efficiency of 32% with the wall-plug power of 170 MW.

## 1 INTRODUCTION

As a part of the quest for ever-increasing energy for higher energy reactions, an e+e- linear collider at a few hundred GeV to a TeV C.M. energy has been proposed by several high-energy accelerator laboratories in the world [1-2]. There are many general requirements for such colliders, such as high reliability and availability, reasonable power efficiency, lower construction cost, simplicity, easy, and flexible operation. All the requirements are pursued as global guidelines and boundary conditions for such kinds of large-scale machines. The conventional technology for klystron modulators is not suitable for meeting most of the above requirements. A smart modulator is necessary to realize a linear collider with a reasonable performance. A prototype C-band modulator has been developed to establish the technologies for a smart modulator. The reliable performance of the smart modulator was confirmed, and the operating characteristics were analyzed. We describe the detailed analyses of the power consumption and the efficiency of a prototype C-band RF system. Effective ways to increase the RF system efficiency are proposed, and the corresponding reduction in the AC power consumption is evaluated for an e+e- linear collider at a 500-GeV C.M. energy.

## 2 SMART MODULATOR

The C-band scheme has 4,080 klystrons (50-MW peak) and modulators (111-MW peak) for two main linacs [3]. An accelerating gradient of 30 MV/m, including beam loading, can be generated by 50-MW C-band klystrons in combination with an RF compression system. Two klystrons are powered by two independent pulse modulators. The peak RF power of 350 MW generated by the pulse compressor is fed to four accelerating structures. One unit can increase the beam energy to about 120 MeV.

The matching modulator for a C-band klystron has to produce pulses with 2.5-µs flat tops. Due to the rise and fall times, it must produce at least a 4.0-µs pulse with 111-MW peak power. A peak voltage of 350 kV and a peak current of 317 A are required. The nominal load impedance at the peak voltage is 1104 Ω. The energy per pulse is 389 Joules. The average power for operation with 100-pps pulses is about 40 kW.

A smart modulator satisfying the RAM (reliability + availability + maintainability) requirements should have a MTBF (mean time between failures) of over 35,000 hours and a MTTR (mean time to repair) of less than 25 minutes in order to realize large-scale linear colliders. The smart modulator also has a compact footprint with a density that is at least two times higher than that of a conventional modulator.

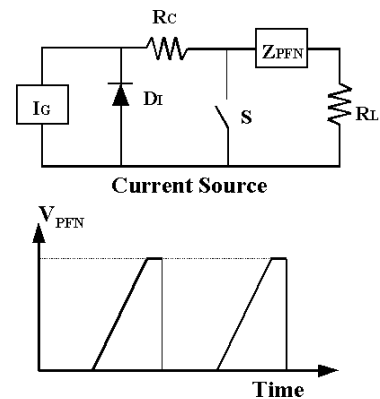

Figure 1 : Block diagram of the smart modulator.

A smart modulator is different from the conventional modulator for the charging power supply, as shown in the Fig. 1. The charging power supply is a constant current source by an inverter power supply. The short circuit current is limited less than a few amperes by the inverter power supply. The PFN charging starts after a long delay, which is controlled by a master trigger-generator, which will certainly provide fault-free operation for the thyratron.

Figure 2 is a simplified circuit diagram of a modulator. We do not need a De-Q'ing system to regulate the PFN voltage. The PFN is an 18-stage Guillemin E-type LC network. There is no bulky thyrite to present the low inverse voltage required for thyratron deionization in the EOLC circuit due to the command-charging scheme of the inverter power supply.

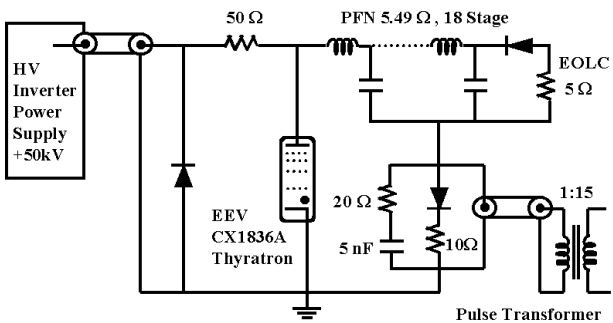

Figure 2 : Circuit schematic of the C-band modulator.

## 3 PERFORMANCE OF MODULATOR

### 3.1 Pulse Characteristics

Figure 3 shows the pulse waveform on the klystron load. The leading edge of the pulse is clean due to low noise. The peak voltage at the pulse flat top is 350 kV and is measured using a capacitive voltage divider mounted inside the pulse tank. The pulse width measured at 75% of the peak value is 4.35 µs. The rise time (10-90%) is 0.96 µs. The fall time is 1.8 µs. A flat top width of 2.5 µs with a ripple of less than ±0.5% is achieved after fine tuning the PFN. A pulse stability of 0.35% is mainly due to the regulation precision of the inverter power supply. The corresponding phase variation of the klystron RF output is about 2°.

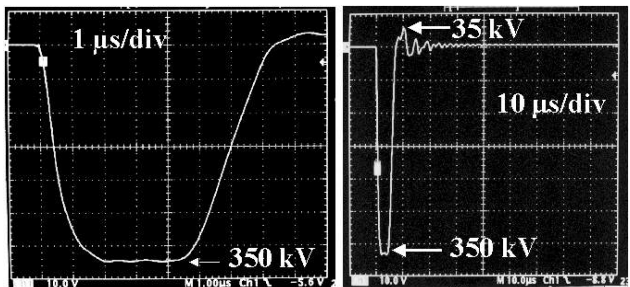

Figure 3 : Klystron pulse waveform.

The peak inverse voltage is less than 35 kV that are caused by the energy stored in the leakage inductance of the pulse transformer. The backswing amplitude is less than 13 kV and is controlled by the resistance of the tail clipper and the magnetizing current of the pulse transformer. The inverse voltage is quite small to guarantee reliable operation of the active components, such as the inverter power supply, the thyratron, and the klystron tube. The measured time jitter of the klystron beam pulses at a 350-kV pulse voltage is 1.2 ns. The total accumulated acquisition time is 11.4 hours, which means that this jitter is the fluctuation of the time delay for $2 \times 10^6$ shots.

### 3.2 Inverter Characteristics

A flexible command-charging scheme is essential when using an inverter power supply. Figure 4 shows charging and discharging waveform of the PFN. The inverter power supply (EMI, model 303) starts to charge the PFN at an arbitrary time. The PFN is linearly charged up to 48 kV within 14 ms. After discharging the PFN, the next charging is inhibited. The charging slope has a fine structure that has a 56-kHz pattern. Each 62-V step is the result of one switching cycle of the inverter power supply. This is the minimum controllable magnitude obtainable when using an inverter power supply. The stability of the beam voltage is directly affected by the fluctuation, which is 0.39% at a charging level of 48 kV without an additional de-Q'ing module.

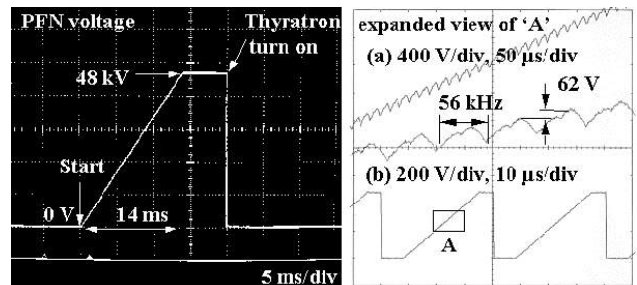

Figure 4 : Charging and discharging of the PFN.

## 4 POWER DISTRIBUTION ANALYSIS

The power distribution for one unit of a C-band pulse modulator is described in Table 1. This analysis is based on an operating condition with full-power specifications, which means a peak voltage of 350 kV and a pulse repetition rate of 100 Hz. A beam power of 15.3 kW is dumped to the klystron collector within the flat-top portion. This loss caused by the conversion efficiency of the klystron is 23% of the total power. The charging loss is due mainly to the switching loss of the IGBT and to the eddy-current loss of the high-frequency transformer in the inverter power supply.

The wasted pulse power includes all losses that do not contribute to generating RF power from the energy stored in the PFN. The loss during the rise and the fall times is 19.4 kW and includes both the energies stored in the leakage inductance and in the distributed capacitance, and the electron beam power delivered to the collector during this time interval. The loss during the rise and fall times is over 90% of the total wasted pulse power. Therefore, it is the main loss in a pulse modulator, and accounts for 29% of the total power. The magnetizing loss of 0.9 kW is due to the energy stored in the shunt inductance of the pulse transformer, which is finally dissipated at the tail clipper.

The solenoid-coil power of the klystron magnet accounts for most of the auxiliary power. The solenoid power of 6.5 kW takes 80% of the total auxiliary power of the C-band system. An efficient way to reduce the auxiliary power is to develop a klystron tube for which the magnetic field is provided by a PPM (periodic permanent magnet) instead of an electric solenoid coil.

Table 1. Power distribution of a C-band system.

| Total Power Consumption = 66.1 kW | |
|---|---|
| 1. RF power | 12.5 kW |
| 2. Wasted beam power | 15.3 kW |
| 3. Wasted pulse power | 21.5 kW |
| 4. Charging loss | 8.7 kW |
| 5. Aux. power | 8.1 kW |
| Details of Wasted Pulse Power | |
| 1. Rise/fall time loss | 19.4 kW |
| 2. Magnetizing loss | 0.9 kW |
| 3. Thyratron loss | 0.4 kW |
| 4. Eddy current loss | 0.5 kW |
| 5. RC snubber loss | 0.3 kW |
| Details of Auxiliary Power | |
| 1. Klystron magnet | 6.5 kW |
| 2. Thyratron heater | 0.6 kW |
| 3. Klystron heater | 0.4 kW |
| 4. Others | 0.6 kW |

## 5  EFFICIENCY EVALUATION

Table 2 shows the present performance of the C-band RF system, including anticipated future parameters based on the technological improvements for the klystron amplifier and a somewhat larger acceptable ripple. The pulse efficiency of 56.4% is mainly limited by the distributed capacitance in the oil-filled pulse tank. Since the net RF system efficiency is 18.9%, the total AC wall-plug power of 270 MW will be consumed to generate a total RF power of 51 MW when using a 4,080-unit klystron-modulator system. The klystron conversion efficiency is given by the initial baseline design of 45%. It is the most effective parameter for the AC wall-plug power, so intensive research to get a higher klystron conversion efficiency is necessary. The pulsing efficiency is almost optimized and is fixed by the reliability criteria for the pulse-transformer design. Relaxation of the ripple requirement, which can be achieved by using a pulse compressor and PM-AM modulation, will allow a larger ripple of up to 2-3%; then, a higher pulse efficiency is expected [4]. The charging efficiency can be increased by a trade-off with a new design for a specific system. Then, C-band main linacs can be operated with an AC wall-plug power of less than 170 MW.

Table 2. System efficiency and power consumption.

| Parameters | Present System | Upgrade System |
|---|---|---|
| Charging Efficiency (%) | 85 | 90 |
| Pulsing Efficiency (%) | 56.4 | 75 |
| Modulator Efficiency (%) | 47.9 | 67.5 |
| Klystron Efficiency (%) | 45 | 60 |
| K-M Efficiency (%) | 21.6 | 40.5 |
| RF System Efficiency (%) | 18.9 | 32 |
| AC Power/One Unit (kW) | 66.1 | 40.8 |
| Total RF Power (MW) | 51 | 51 |
| Total AC Power (MW) | 270 | 170 |

## 6  FUTURE R&D ITEMS

In conventional modulator systems, the thyratron is the most troublesome component. It has a relatively short lifetime (approximately $3 \times 10^9$ shots), which varies from tube to tube, and installation procedures which are not simple. The protection circuits need substantial space since they are currently assembled by using a large number of diodes, snubber capacitors, and resistors. The conventional modulator tank is expensive and massive; it also needs insulation oil. The voltage monitor is less reliable. Most of the final design has been passively determined, resulting in a conservative space margin after finishing the component design and fabrication. Although the inverter power supply can provide essential functions in a smart modulator, the stability of the inverter module itself is still not satisfactory. One possible way to improve the power efficiency is to recover the tail energy of the HV pulse by storing energy in the buffer capacitor.